\documentclass[11pt,twoside]{article}
\usepackage[pdftex]{graphicx}
\usepackage{amsmath}
\usepackage{amssymb}
\usepackage{cite}

\begin{document}
\newcommand{\pst}{\hspace*{1.5em}}

\newcommand{\rigmark}{\em Journal of Russian Laser Research}
\newcommand{\lemark}{\em Volume 35, Number 1, 2014}

\newcommand{\be}{\begin{equation}}
\newcommand{\ee}{\end{equation}}
\newcommand{\bm}{\boldmath}
\newcommand{\ds}{\displaystyle}
\newcommand{\bea}{\begin{eqnarray}}
\newcommand{\eea}{\end{eqnarray}}
\newcommand{\ba}{\begin{array}}
\newcommand{\ea}{\end{array}}
\newcommand{\arcsinh}{\mathop{\rm arcsinh}\nolimits}
\newcommand{\arctanh}{\mathop{\rm arctanh}\nolimits}
\newcommand{\bc}{\begin{center}}
\newcommand{\ec}{\end{center}}

\thispagestyle{plain}

\label{sh}


\begin{center} {\Large \bf
\begin{tabular}{c}
Balance equations-based properties of the Rabi Hamiltonian
\end{tabular}
 } \end{center}

\bigskip

\bigskip

\begin{center} {\bf
Antonino Messina,$^{1\,*}$ Anna Napoli,$^1$ Margarita A. Man'ko,$^2$ and
Vladimir I. Man'ko$^2$ }\end{center}

\medskip

\begin{center}
{\it $^1$Dipartimento di Fisica e Chimica, University of Palermo\\ Via
Archirafi 36, Palermo 90123, Italy }

\smallskip

{\it $^2$P.~N.~Lebedev Physical Institute, Russian Academy of Sciences\\
Leninskii Prospect 53, Moscow 119991, Russia}

\smallskip

\smallskip

$^*$Corresponding author e-mail:~~~antonino.messina\,@\,unipa.it
\end{center}

\begin{abstract}\noindent
A stationary physical system satisfies peculiar balance conditions involving mean values of appropriate observables.  In this paper we show how to deduce such quantitative links, named balance equations, demonstrating as well their usefulness in bringing to light physical properties of the system without solving the Schrodinger equation.  The knowledge of such properties in the case of Rabi Hamiltonian is exploit to provide arguments to make easier the variational engineering of the ground state of this model. 
\end{abstract}

\medskip

\noindent{\bf Keywords:} ground state, variational approach, Wigner function,
Rabi model, balance equations.

\section{Introduction}
\pst Let $H(\eta)$ be the time-independent Hamiltonian of a binary physical
system $S$, whose states $\{ \mid{\phi}\rangle\}$ live and evolve in the
Hilbert space $\mathcal{H}$. We assume that such a Hermitian operator depends
on a real (set of) parameter(s) $\eta$. We denote by
$\mid{\psi_E(\eta)}\rangle$ a normalized eigenvector of $H(\eta)$  of eigenvalue $E(\eta)$. We recall that in the Schr\"odinger
representation the time derivative of an operator $A$ (generally depending
both on $\eta$ and explicitly on $t$), denoted by $\dfrac{dA}{dt}$, is
defined as ($\hbar=1$)
\begin{equation}\label{def_der_A}
\langle{\phi(t)}\mid\frac{d}{dt}A\mid{\phi(t)}\rangle\equiv
\frac{d}{dt}\langle{\phi(0)}\mid e^{iH t}Ae^{-iH t}\mid{\phi(0)}\rangle .
\end{equation}
It is well known that such a definition leads to
\begin{equation}\label{eq_der_A}
\frac{d}{dt}A=\frac{\partial}{\partial t}A +i[H,A],
\end{equation}
which, on the one hand, implies that $A$ is a constant of motion if either
$[A,H]=0$ or the partial time-derivative contribution is compensated by the
commutator term (see, e.g., \cite{183}). On the other hand,
Eqs.~(\ref{def_der_A}) and (\ref{eq_der_A}) guarantee that the expectation
value of any time-independent operator $A$ taken on any stationary state
$\mid{\psi_E(\eta)}\rangle$ of $H(\eta)$ vanishes. Since, in general,
$A$ does not commute with $H(\eta)$, Eq.~(\ref{eq_der_A}) originates in
each stationary state a link between the mean values of the operators
additively stemming from the commutator $[A,H]$. This algebraic relation is an
identity with respect to $\eta$ and, when $A$ is an observable, it reflects, in any
stationary state of $H(\eta)$, the existence of a necessary
quantitative relationship among the $\eta$-dependent expectation values of
appropriate operators which, being Hermitian like $i[A,H]$, are amenable to
a physical interpretation.

Following \cite{Messina1,Messina2,Messina3,Messina4,Messina5}, we call these
equations balance equations associated to $H(\eta)$.

The interest toward the balance equations is at least threefold.

First of all, they are the fruit of a simple calculation and might help to
highlight physical properties common to all the stationary states of $S$, in
particular, its ground state, without solving the time-independent
Schr\"odinger equation which might indeed be difficult to handle with. 
On the other hand, considered that when this is the case one generally looks for approximate expressions of the eigensolutions of the hamiltonian model under scrutiny, the balance equations might provide an exact tool to check, a posteriori, the accuracy of the approximated solution found. As third and final remark, we in addition emphasize that the knowledge of a set of easily constructed and physically transparent balance equations associated to a given hamiltonian, might put at our disposal arguments/constrains to control a priori the quality of the approximation route. For example we might be addressed, on physical grounds, to express $H(\eta)$ in a selected basis where it might be diagonalized with the help of a convincing truncation protocol or, within a variational approach aimed at finding the ground state of the system under study, we might wisely guess a''lucky" class of trial states.


In this paper, we highlight the usefulness and advantages of such balance equations-based  approach
by considering,  as  application, the variational engineering of the ground state of the following ever appealing and always on the fashion quantum
Rabi model  \cite{Rabi1, Rabi2, Rabi3}
\begin{equation}\label{Rabi model}
H=\omega\alpha^{\dag}\alpha
+\lambda (\alpha+\alpha^{\dag})\sigma_x+\frac{\omega_0}{2}\sigma_z\equiv H(\eta)
\end{equation}
describing the linear coupling of strength $\lambda$ between a quantum
harmonic oscillator (or a single bosonic mode) of angular frequency $\omega$
and an effective two-level system (or a spin ${1}/{2}$) with the Bohr
frequency $\omega_0$. The dynamical variables of the quantum mode are the
annihilation and creation operators $\alpha$ and $\alpha^{\dag}$ whereas the two-level system is completely represented by the
Pauli matrices $\sigma_x$, $\sigma_y$, and $\sigma_z$.

Over the last 70 years this paradigmatic model
has been 
investigated in a myriad of
papers, even in its multimode version useful to treat the two-level system as
an open quantum system \cite{Breuer}. Many facets of its static, dynamical, and
thermodynamical behavior have been theoretically disclosed and experimentally
revealed in a lot of quite different physical contexts as, for example, 
cavity, circuit, solid state quantum electrodynamics, quantum information and so on \cite{Haroche,2,3,4,5,6}. Quite recently advancements on the exact analytical representation of the eigensolutions of $H$, motivated by the experimental realization of ultra strong coupling regimes, have been reported \cite{Rabi2, Moroz,Gunter,Todorov}.

The Rabi model depends on two independent effective real parameters defining a
bidimensional space $S$ which contains a region of experimental interest where,
however, the physical behavior of the system is analytically less
characterized with respect to that the system exhibits in the complementary
region of its parameter space. In this challenging region, the relative weight
of the three parameters $\omega$, $\lambda$, and $\omega_0$ does not
legitimate any obvious perturbative treatment of $H$, so that the entanglement
get established between the two parties in such a condition reflects the
occurrence of a mutual influence higher than that exhibited by the system out
of this region. Such a situation singles out the so-called intermediate
coupling regime between the weak and the strong regimes realized by the system
when ${\lambda^2}\ll {\omega \omega_0}$ and ${\lambda^2} \gg
{\omega \omega_0}$, respectively.

We stress from the very beginning that we do not intend here to improve the quality of the variational ground state of the Rabi model as reported in the literature. Rather we wish to provide a concrete example of how the knowledge of an appropriate set of balance equations associated to the Rabi hamiltonian, allows to understand the failure of an optimized specific trial state (for example coherent state) in some regions of the parameter space and at the same time how to improve the class of trial state in order to get a new optimized solution closer to the exact ground state in a larger region of $S$. In other words, the balance equation-based approach, exploited in the variational framework, might provide arguments useful to justify a specific choice of the trial state in accordance with our expectations (that is disposable balance equations) concerning the exact ground state. 

The paper is organized as follows. Section 2 reports some useful general properties possessed by the Rabi system in its ground state while the construction of a set of exact balance equations is presented in the subsequent section. The knowledge of these exact constrains is exploited in section 4 to engineer a class of trial variational ground states and to find out an analytical optimized expression of the ground state. Some conclusive remarks are pointed out in the last section where possible developments based on the novel approach reported in this paper are briefly discussed.


\section{Some General Properties of the Rabi Hamiltonian Ground State}
\pst The derivation of balance equations associated to the Rabi hamiltonian is postponed to the next section. Here, instead, we wish to resume and/or to derive some exact properties of the ground state of this model. Such properties, conjugated with the balance equations, play an interesting role since they reveal in a transparent way the nature of $\eta$-dependent constrains in the structure of the exact ground state of the Rabi model, which then must be taken into account when tailoring the analytical form of a variational trial state.   It is easy to prove that the Hermitian and unitary parity operator
$P=-\sigma_z \cos(\pi\alpha^{\dag}\alpha)$ commutes with $H$, so that the normalized
stationary states  $\mid{\psi_{E,\,p}}\rangle$ of $H$ of definite parity $p=\pm 1$, belonging to the energy eigenvalue $E$, may be represented as
follows:
\begin{equation}
\mid{\psi_{E,\,p}}\rangle=\sum_{n=0}^\infty
a_n^{(p)}\mid{n}\rangle\mid{\sigma=(-1)^{n+1}p}\rangle
\end{equation}
provided that the bosonic state
\begin{equation}\label{2.1}
\mid{\phi_{E,\,p}}\rangle=\sum_{n=0}^\infty a_n^{(p)}\mid{n}\rangle
\end{equation}
is normalized.

Since the knowledge of $\mid{\phi_{E,\,p}}\rangle$ univocally determines
$\mid{\psi_{E,\,p}}\rangle$, it is not surprising that, as a consequence of the peculiar spin--boson entanglement induced
by $P$, the Rabi Hamiltonian
may be unitarily traced back to the following $p$-dependent bosonic
Hamiltonian
\begin{equation}\label{h tilde}
\tilde{H}_p=\omega \alpha^{\dag}\alpha +\lambda
(\alpha+\alpha^{\dag})-\frac{\omega_0}{2}p \cos\,(\pi \alpha^{\dag}\alpha),
\end{equation}
where $p=\pm 1$ is an eigenvalue of $P$.

It has been rigorously demonstrated \cite{Messina4,Messina5} that the ground state of $\tilde{H}_+$
generates the ground state $\mid{g_{+1}}\rangle$ of $H$ everywhere in $S$, and that
there exist regions of $S$ where the ground state of $\tilde{H}_{-1}$ is
degenerate with that found in $\tilde{H}_+$. This happens, for example, when
$\omega_0=0$ and, in such a case,  ground states of $H$ of no definite
parity exist. It is easy to convince oneself that the probability amplitudes
of the ground state of $\tilde{H}_p$ (and then of $H$) may be chosen all real
without loss of generality.

It is useful to write  $\mid{\psi_{E,\,p}}\rangle$ exploiting the eigenstates
of $\sigma_x$ instead of those of $\sigma_z$. It is not difficult to show that
\begin{equation} \label{psi E p}
\mid{\Psi_{E,\,p}}\rangle=\frac{1}{\sqrt{2}}\big\{\mid{\phi_{E,\,p}}\rangle\mid{+}\rangle_x-
p \cos\,(\pi \alpha^{\dag}\alpha)\mid{\phi_{E,\,p}}\rangle\mid{-}\rangle_x
\big\}.
\end{equation}

Let $E_g$ be the exact ground state energy of the Rabi model. Since,
occasionally, $E_g$ may result degenerate, we simply denote by
$\mid{g_p}\rangle$ a (the) solution of $H\mid{g}\rangle=E_g\mid{g}\rangle$
having a definite parity $p$. It is possible to show the validity of the
following properties everywhere in $S$: 
\begin{eqnarray}
-\frac{\omega_0}{2}-\frac{\lambda^2}{\omega}\leq E_g\leq -\frac{\omega_0}{2},
\label{p1}\\
\langle {g_p}\mid\sigma_z\mid{g_p}\rangle=-p\langle{g_p}\mid\cos\,(\pi \alpha^{\dag}\alpha)\mid{g_p}\rangle \leq 0, \label{p2}\\
\langle{g_p}\mid(\alpha+\alpha^{\dag})\sigma_x\mid{g_p}\rangle \leq 0, \label{p3}\\
-\omega_0\leq\omega \langle {g_p}\mid \alpha^{\dag}\alpha\cos\,(\pi
\alpha^{\dag}\alpha)\mid{g_p}\rangle=-\omega
\langle{g_p}\mid\alpha^{\dag}\alpha \sigma_z\mid{g_p}\rangle \leq \omega_0.
\label{p4}
\end{eqnarray}
Equations~(\ref{p1})--(\ref{p4}) are certainly valid for $\mid{g_{+1}}\rangle$
(the ground state of $H$ of parity $p=+1$ exists in all points of $S$) and
also for $\mid{g_{-1}}\rangle$ in the  case of degeneration.
Equation~(\ref{p1}) stems from elementary considerations based on the position
of $E_g$ in the energy spectrum of $H$. Equation~(\ref{p2}) reflects the
property $P\mid{g_p}\rangle=p\mid{g_p}\rangle$ (then valid also outside the
minimum energy subspace), as well as that the probability of finding the
oscillator in its ground state $\mid{g_{+1}}\rangle$ ($\mid{g_{-1}}\rangle$)
with an even number of excitations exceeds (is less than) that of finding the
oscillator with an odd number of excitations. Equation~(\ref{p3}) means that
the covariance of the adimensional coordinate of the quantum oscillator and the
``coordinate'' of the two-level system is always negative in $S$, since
$\langle{g_p}\mid(\alpha+\alpha^{\dag})\mid{g_p}\rangle=\langle{g_p}\mid\sigma_x\mid{g_p}\rangle=0$
for symmetry reasons. Moreover, this equation says that the interaction energy
counters the nonnegative contribution of the free energy of the quantum
oscillator on $\mid{g_p}\rangle$ in accordance with the requirement for $E_g$
prescribed by Eq.~(\ref{p1}). This last comment is also valid for the
expectation value of $\sigma_z$ on $\mid{g_p}\rangle$.

Equations~(\ref{p4}) and (\ref{p2}) reveal the occurrence of a limited
variability for $\left|\langle{g_p}\mid(\alpha^{\dag}\alpha)
x\mid{g_p}\rangle\right|$, with $x= \cos\,(\pi \alpha^{\dag}\alpha)$ or
$x=\sigma_z$, traceable back to the negativity of $E_g$ everywhere in $S$. The
link between $ \langle{g_p}\mid \alpha^{\dag}\alpha\cos\,(\pi
\alpha^{\dag}\alpha)\mid{g_p}\rangle$ and $\langle{g_p}\mid\alpha^{\dag}\alpha
\sigma_z\mid{g_p}\rangle$ is once again consequence of the entanglement
introduced in the binary Rabi system by the parity constraint.
In principle, other identities and inequalities in $S$, as done for
Eqs.~(\ref{p1}) and (\ref{p4}), may be systematically constructed exploiting
Eq.~(\ref{2.1}) and the relations deducible by taking the mean value in the
ground state $\mid{g_p}\rangle$ of the operator equation expressing the
anticommutator between $H$ and appropriate observables. From the mathematical
point of view, the only hypothesis used to derive Eqs.~(\ref{p1})--(\ref{p4})
is the assumed existence of all the expectation values involved, since, in accordance with our procedure, we are
not making use of the analytical form of $\mid{g_p}\rangle$.

\section{Balance Equations}
\pst In this section, we demonstrate that all the stationary states of $H$
share the occurrence of analytical links valid everywhere in $S$ among the expectation
values of selected and physically interpretable observables as anticipated in the introduction. We call such
relations balance equations even if, strictly speaking, they are
identities in $S$. The point making the balance equations of theoretical
interest consists in the fact that they may be systematically generated as
necessary conditions of the stationarity, without any a priori knowledge of
the analytical form of the eigensolutions of $H$.

To make easier capturing the physical meaning of the results to be derived in
this section, we write the Rabi hamiltonian describing the bosonic mode as
that for a quantum oscillator in its phase space
\begin{equation}\label{oscillator h}
H=\frac{p^2}{2m}+\frac{1}{2} m\omega_2 q^2+F_0q\sigma_x+\frac{\omega_0}{2}
\sigma_z-\frac{\omega}{2}\,,
\end{equation}
where
\begin{equation}
q=\sqrt{\frac{1}{2m\omega}}(\alpha+\alpha^{\dag}),\qquad 
p=i \sqrt{\frac{m\omega}{2}}(\alpha^{\dag}-\alpha),\qquad 
F_0=\sqrt{2m\omega}\lambda. \label{F0}
\end{equation}
Since, in view of Eq.~(\ref{eq_der_A}),
\begin{equation}\label{eq F}
m\frac{d^2 q}{dt^2}=-m\omega^2q-F_0\sigma_x\equiv F_q+F_e,
\end{equation}
we immediately derive 
\begin{equation}
\langle {\psi_E(\eta)}\mid F_q\mid{\psi_E(\eta)}\rangle=-\langle {\psi_E(\eta)}\mid F_e\mid{\psi_E(\eta)}\rangle.
\end{equation}

Thus 
in the ground state of the Rabi hamiltonian the
elastic force $F_q$ and external force $F_e$ due to the two-level subsystem are, on the
average, opposite. When the ground state has a definite parity both vanish and eq.~(\ref{p3}) tell us that the two forces are anticorrelated since the
expectation values of $q$ and $\sigma_x$ vanish on $\mid{g_p}\rangle$.

To further appreciate how a balance equation may contribute to bring light to
peculiar properties of the ground state of the Rabi system, we exploit the
fact that the mean value on $\mid{\Psi_{E,p}}\rangle$ of the operator
$\dfrac{d^2(q\sigma_x)}{dt^2}$ vanishes.  The resulting balance equation on $\mid{g_p}\rangle$ in particular becomes
\begin{equation}\label{b1}
\langle{g_p}\mid {p^2}/{2m}\mid{g_p}\rangle= \frac{F_0}{2}\langle{g_p}\mid q
\sigma_x\mid{g_p}\rangle+\langle{g_p}\mid m\omega^2q^2/2\mid{g_p}\rangle,
\end{equation}
where the Fock states in the expression of $\mid{g_p}\rangle$ are to be
thought in the $q$-representation.

Getting rid of $\langle{g_p}\mid {p^2}/{2m}\mid{g_p}\rangle$ between
Eq.~(\ref{b1}) and the expression of $E_g$ formally deducible from
Eq.~(\ref{oscillator h}), the resulting equation from such elimination may be
used to cast the double limitation on the lowest energy eigenvalue of $H$, as
given by Eq.~(\ref{eq F}), in the following form:
\begin{equation}\label{b2}
\frac{1}{2m\omega}-\frac{\lambda^2}{m\omega^3}+C\leq\bigtriangleup^2(q\sigma_x)
\leq\frac{1}{2m\omega}+C,
\end{equation}
where $\bigtriangleup^2(q\sigma_x)\equiv \langle{g_p}\mid
(q\sigma_x)^2\mid{g_p}\rangle-\langle{g_p}\mid q\sigma_x\mid{g_p}\rangle^2$
and
\begin{equation}\label{b3}
m\omega^2C=-(1+\langle{g_p}\mid\sigma_z\mid{g_p}\rangle)-3F_0\langle{g_p}\mid
q\sigma_x\mid{g_p}\rangle-\langle{g_p}\mid q\sigma_x\mid{g_p}\rangle^2.
\end{equation}

Considering that, in view of Eq.~(\ref{psi E p}), $\mid{g_p}\rangle$ may be
formally written as
\begin{equation}\label{b4}
\mid{g_p}\rangle=\frac{1}{\sqrt{2}}\big\{\mid{\phi_{E_g,\,p}}\rangle\mid
{+}\rangle_x - p \cos\,(\pi
\alpha^{\dag}\alpha)\mid{\phi_{E_g,\,p}}\rangle\mid{-}\rangle_x \big\},
\end{equation}
we immediately see that
\begin{equation}\label{b5}
\langle{g_p}\mid q\sigma_x\mid{g_p}\rangle=\langle{\phi_{E_g,\,p}}\mid q\mid
{\phi_{E_g,\,p}}\rangle,
\end{equation}
and then
\begin{equation}\label{b6}
\bigtriangleup^2(q\sigma_x)=\bigtriangleup^2_\phi(q),\quad\mbox{where}\quad
\bigtriangleup^2_\phi(q)=\langle{\phi_{E_g,\,p}}\mid
q^2\mid{\phi_{E_g,\,p}}\rangle-\langle{\phi_{E_g,\,p}}\mid
q\mid{\phi_{E_g,\,p}}\rangle.
\end{equation}

Since the fluctuations of $q$ in a coherent bosonic state 
is ${1}/{2m\omega}$, independent on its amplitude, the exact
inequality~(\ref{b2}) suggests that the fluctuation of $q$ on
$\mid{\phi_{E_g,p}}\rangle$ might exhibit values different from
${1}/{2m\omega}$ in selected domains of the parameter space $S$. This observation
is of relevance since it explains why an optimized coherent state fails in representing $\mid{\psi_{g,+1}}\rangle$ everywhere in $S$, as indeed found in the literature without however any attempt to go beyond \cite{Messina1,Messina2}. Thus, we are
interested in finding the argument strengthening the hypothesis that to
overcome such a failure we must introduce a variational trial  state exhibiting
flexibility in the fluctuation of $q$ in different coupling regimes.

To this end, we now deduce another balance equation based on the second
derivative of $\omega \alpha^\dag\alpha$. The final result may be expressed in
the following suggestive form when restricted to the ground state
$\mid{g_p}\rangle$
\begin{equation}\label{b7}
\langle{g_p}\mid F_qF_e\mid {g_p}\rangle+\langle {g_p}\mid p
\,\frac{dF_e}{dt}\,\mid{g_p}\rangle +F^2_0=0,\quad\mbox{with}\quad
\frac{dF_e}{dt}=F_0\omega_0\sigma_y.
\end{equation}
This balance equation discloses the existence in each point of $S$ of a link
between the covariance of the two forces $F_q$ and $F_e$ on the oscillator and
the covariance between the oscillator momentum and the rapidity of variation
of $F_e$. In particular, it says that coupling regimes, where the correlations
between the two forces are almost vanishing, are characterized by an
anticorrelation between $p$ and $\dfrac{dF_e}{dt}$. The relevance of this
comment may be elucidated by the consideration that for $\omega_0<\omega$, on
the one hand, whatever $\eta$ is, $\mid{\phi_{E_g,\,+1}}\rangle$ may be
well approximated by an appropriate coherent state exhibiting an effective
displacement $\bar{q}$.

When instead  $\omega_0\gg\omega$, in the region of $S$ where
$\lambda^2\ll\omega_0\omega$, $F_q$ and $F_e$ decorrelate more and more
approaching zero when ${\lambda^2}/{\omega_0\omega}$ tends to zero. In such a
condition, the correlation between $p$ and $\dfrac{dF_e}{dt}$, in view of
Eq.~(\ref{b7}), grows in absolute value approaching its minimum negative value
$-F_0^2$. Thus, in this region of $S$, the anticorrelation between $p$ and
$\dfrac{dF_e}{dt}$ provokes the diminution of the displacement of the
oscillator with respect to $\bar{q}$. This means that, when
$\omega_0\gg\omega$ and $\lambda$ is such to guarantee a weak coupling regime,
the ground state of the Rabi Hamiltonian exhibits an almost vanishing mean
value of $q$ but the spread of the same observable is greater than that
associated to the coherent state occurring when $\omega_0<\omega$.
Mathematically this fact stems from the inevitable presence in
$\mid{\phi_{E_g,\,+1}}\rangle$ of the odd Fock states necessary to comply with
the condition  $\langle{g_p}\mid p \dfrac{dF_e}{dt}\mid{g_p}\rangle<0$.

This heuristic analysis is qualitatively compatible with the double
inequality~(\ref{b2}) and provides an example of a region of $S$, where
certainly the trial choice of  $\mid{\phi_{E_g,\,+1}}\rangle$ in the form of
the coherent state is not legitimate. On the basis of the suggestions stemming
from the arguments developed in this section, we can construct a proposal for
$\mid{g_{+1}}\rangle$ flexible enough to comply with all the balance
equations, exact necessary conditions, and to recover its coherent state-based
description when $\omega_0<\omega$.

\section{Engineering a Variational Ground State}
\pst On the basis of what we have learned  and highlighted about the properties of the ground
state of parity $+1$ in $S$, we must go beyond the trial choice of
$\mid{\Phi_{E_g,\,+1}}\rangle$,  as given by eq. ~(\ref{2.1}), since the coherent state $D(\beta)\mid{0}\rangle$ of optimizable amplitude $\beta$
essentially becomes incompatible with the double inequality~(\ref{b2})
everywhere in $S$.
The operator $D(\beta)$ is the unitary displacement operator defined as
\begin{equation}\label{4.3}
D(\beta)=\exp\{\beta\alpha^\dag -\beta^*\alpha \}.
\end{equation}

To gain more flexibility in the fluctuations of
$(\alpha+\alpha^\dag)$ not renouncing as well to a coherent recover of
$\mid{\phi_{E_g,\,+1}}\rangle$, where appropriate in $S$, we propose the
following two real-parameter squeezed and displaced states:
\begin{equation}\label{4.1}
\mid{\phi_{E_g,\,+1}(\beta,\gamma)}\rangle=S(\gamma)D(\beta)\mid{0}\rangle,
\end{equation}
the squeezing unitary operator $S(\gamma)$ being
\begin{equation}\label{4.2}
S(\gamma)=\exp\big\{{\gamma}\big(\alpha^{\dag 2}-\alpha^2\big)/2\big\}.
\end{equation}
One can demonstrate that  $\mid{\phi_{E_g,\,+1}(\beta,\gamma)}\rangle$ identically satisfies
Eqs.~(\ref{p1})--(\ref{p4}) and that the two equations in $\gamma$ and $\beta$
obtained from the balance equations~(\ref{b1}) and (\ref{b7}) coincide with
the two variational equations determined from the optimization of the energy
functional of the system in the class of states~(\ref{4.1}) with respect to
the two variational parameters $\gamma$ and $\beta$. Moreover, Eq.~(\ref{4.1})
is compatible with Eq.~(\ref{b2}) since, when $\gamma=0$, the squeezed
displaced state gives back the coherent state  $\mid{\beta}\rangle$ and, in general, meets all the
requests on the ground state built so far in this paper. To find the optimized  dependence
of $\gamma$ and $\beta$ on the model parameters, we must evaluate the energy
$E(\beta,\gamma)$, that is,
\begin{equation}\label{4.4}
E(\beta,\gamma)=\langle{0}\mid
S^\dag(\gamma)D^\dag(\beta)\tilde{H}_+D(\beta)S(\gamma)\mid{0}\rangle,
\end{equation}
where from Eq.~(\ref{h tilde})
\begin{equation}\label{4.5}
\tilde{H}_+=\omega \alpha^{\dag}\alpha
+\lambda (\alpha+\alpha^{\dag})-\frac{\omega_0}{2} \cos( \pi \alpha^{\dag}\alpha)
\end{equation}
is a nonlinear restriction of $H$ in the parity-invariant subspace with
$p=+1$ where the ground state certainly is.

To this end, in the following we deduce an interesting link between the mean
value $E$ of the Rabi reduced Hamiltonian $\tilde{H}_+$ in any arbitrary pure
state of the quantum bosonic mode and the value $W(0,0)$ assumed by its Wigner
function $W(p,q)$~\cite{Wigner32} in the same state. It is well known that the
expectation value of the parity operator $\cos\,(\pi \alpha^\dag \alpha)$ is
related to
\begin{equation}\label{4.6}
W(0,0)=2\int \langle x\mid \phi\rangle\langle\phi\mid -x\rangle\, dx
\end{equation}
in a generic state $\mid{\phi}\rangle$ given in the Fock representation by
\begin{equation}\label{4.7}
\langle{\phi}\mid \cos\,( \pi \alpha^{\dag}\alpha)
\mid{\phi}\rangle=W(0,0)/2=\mbox{Tr}\,\big(\rho \cos\,( \pi
\alpha^{\dag}\alpha)\big ),
\end{equation}
where $\rho=\mid{\phi}\rangle\langle{\phi}\mid$. Then
\begin{eqnarray}\label{4.8}
E&\equiv&\mbox{Tr}\,(\rho \tilde{H}_+)=\mbox{Tr}\,\big(\rho[\omega
\alpha^{\dag}\alpha +\lambda (\alpha+\alpha^{\dag})]\big)-\mbox{Tr}\,\big(\rho
[{\omega_0}/{2}] \cos\,( \pi \alpha^{\dag}\alpha)\big)\nonumber \\
&=&\mbox{Tr}\,\big(\tilde{\rho}[\omega \alpha^{\dag}\alpha
-{\lambda^2}/{2\omega}]\big)-({\omega_0}/{4})W(0,0) =
\omega\langle\tilde{n}\rangle-({\lambda^2}/{2\omega})-({\omega_0}/{4})W(0,0),
\end{eqnarray}
where $\tilde{\rho}=D^\dag(-{\lambda}/{\omega})\rho D(-{\lambda}/{\omega})$,
$D(-{\lambda}/{\omega})$ being the displacement operator accomplishing
the exact diagonalization of $\tilde{H}_+$ when $\omega_0=0$. The mean value
of $\alpha^\dag \alpha$ in the state $\tilde{\rho}$ is here denoted by
$\langle\tilde{n}\rangle$. Considering that $|W(q,p)|\leq 2$~\cite{OConnel} in
the oscillator phase space, we immediately arrive at
\begin{equation}\label{4.9}
-\frac{\omega_0}{2}-\frac{2\lambda^2}{\omega}\leq
E(\beta,\gamma)-\omega\langle\tilde{n}\rangle \leq
\frac{\omega_0}{2}-\frac{2\lambda^2}{\omega}\,.
\end{equation}

When $\mid{\phi}\rangle$ belongs to the class of trial states given by
Eq.~(\ref{4.1}), $W(0,0)=e^{-2\beta^2}$, and after evaluating $
\langle\tilde{n}\rangle$, we obtain
\begin{equation}\label{4.10}
E(\beta,\gamma)=\omega\big[\beta^2e^{2\gamma}+\sinh(\gamma^2)\big]+2\lambda\beta
e^\gamma-({\omega_0}/{2})e^{-2\beta^2}.
\end{equation}
Developing this variational approach, we minimize $E(\beta,\gamma)$ finding
$\bar{\beta}$ and $\bar{\gamma}$, variable in $S$, and such that the unitary
operator $S(\bar{\gamma})D(\bar{\beta})$ transforms $\tilde{H}_+$ into the sum
of a diagonal contribution whose ground state is the vacuum state and another
one which may legitimately be considered as perturbative with respect to the
diagonal one. This means that the ground state found within the class of trial
states given by Eq.~(\ref{4.1}) is reasonably close to the exact ground state
and, as a consequence, that we are in condition to investigate properties different from its energy, of
the Rabi system in its ground state, using the variationally optimized fundamental state.

It is interesting to observe that other classes of trial states may be
proposed all fulfilling the balance equations so that a comparative
investigation of their reliability with the class here proposed might lead to
the construction of new more reliable accurate proposals. This task will be faced with in a
successive paper.

\section{Conclusive Remarks}
\pst  In this paper we have presented  and applied the novel idea of balance equation that is a quantitative link existing  among mean values of observables necessary holding in each stationary state of a physical system. Generally speaking the balance equations are infinitely-many and may be seen as a class of constrains making the system stationary. This circumstance has suggested the idea of exploiting the knowledge of even a finite set of such balance equations to introduce a systematic approach to bring to light  physical properties  the system possesses in stationary conditions. To show the concreteness of such a point of view, we have conjugated the balance equations with the variational protocol 
 considering in detail the Rabi model that is one bosonic mode
interacting with one qubit.
For this model, we demonstrated  the usefulness of the balance equations to choose the probe function of its ground
state.
To study the constraints for the
system's ground state energy, we used in particular the known inequalities for the Wigner
functions~\cite{Wigner32,Schleichbook}. Since there exists the probability
representation of quantum states where the wave functions and energies of
stationary states are determined by the tomographic-probability distributions
(see, e.g, \cite{OlgaBeppeJPA} and the recent review~\cite{NuovoCim}) obeying
the corresponding equations~\cite{AmosKorPRA}, the balance-equations approach
can be extended to study the properties of the probability distributions of
the ground state satisfying the quantum equations. Such an extension will be
considered in a future publication.

\section*{Acknowledgments}
\pst M.A.M. and V.I.M. thank the University of Palermo for invitation and kind
hospitality.

\end{document}